\documentclass[pre,aps,groupaddress,showpacs,onecolumn]{revtex4}
\usepackage{epsfig,amsmath,graphicx,amssymb,overpic}

\usepackage{graphicx}% Include figure files
\usepackage{dcolumn}% Align table columns on decimal point
\usepackage{bm}% bold math
\usepackage{graphicx}
\usepackage{subfigure}
\usepackage{amsmath}
\newtheorem{thm}{Theorem}

\usepackage{mathtools}
\usepackage{subfigure}
\usepackage{amsmath}
\usepackage{amssymb}%花体字母加粗
\usepackage{mathrsfs}%花体字母

\def\be{\begin{equation}}
\def\ee{\end{equation}}
\def\bee{\begin{eqnarray}}
\def\ene{\end{eqnarray}}
\def\bes{\begin{subequations}}
\def\ees{\end{subequations}}
\begin{document}

\title{The $\bar{{\partial}}$-dressing Method for Two (2+1)-dimensional Equations and Combinatorics}
\author{Xuedong Chai}\author{Yufeng Zhang}
\email{zyfusf@163.com;zhangyfcumt@163.com }
\affiliation{\vspace{0.1in}
    School of Mathematics, China University of Mining and Technology, Xuzhou, Jiangsu 221116, China \\
    %$^{2}$ School of Mathematical and Statistical Sciences, The University of Texas Rio Grande Valley, Edinburg,TX78539, USA \\
    }
\baselineskip=15pt

%\title{$\bar{{\partial}}$-dressing method for the Sawada Kotera equation}
%\author[label1]{Xuedong Chai$^{1}$}\author[label1]{Yufeng Zhang$^{1}$}
%\author[label2]{Yong Chen$^{2}$} \author[label3]{Shiyin Zhao$^{3}$}
%\email{zhangyfcumt@163.com}
%\affiliation[label1]{\vspace{0.1in}
%    $^{1}$ School of Mathematics, China University of Mining and Technology, Xuzhou, Jiangsu 221116, China \\}
%\affiliation[label2]{\vspace{0.1in}
%    $^{1}$ School of Mathematics, China University of Mining and Technology, Xuzhou, Jiangsu 221116, China \\}
%\baselineskip=15pt

\begin{abstract}\baselineskip=15pt
The soliton solutions of two different (2+1)-dimensional equations with the third order and second order spatial spectral problems are studied by  the  $\bar{{\partial}}$-dressing method.
The 3-reduced  and 5-reduced equations of CKP equation are obtained for the first time by the binary Bell polynomial approach.
The $\bar{{\partial}}$-dressing method is a very important method to explore the solution of a nonlinear soliton equation without analytical regions. %Although there are other ways which could be applied for this aim, but the $\bar{{\partial}}$-dressing method  leads  to the final results directly.
Two different types of equations, the (2+1)-dimensional Kaup-Kuperschmidt equation named CKP equation and a generalized (2+1)-dimensional nonlinear wave equation, are studied by analyzing the eigenfunctions and Green's functions of their Lax representations as well as the inverse spectral transformations, then the new $\bar{{\partial}}$ problems are deduced to construct  soliton solutions  by choosing  proper spectral transformations. Furthermore, once the time evolutions of the spectral data are determined, we will be able to completely obtain the solutions formally of the CKP equation and the generalized (2+1)-dimensional nonlinear wave equation.
The reduced problem of the (2+1)-dimensional Kaup-Kuperschmidt is discussed. The main method is binary Bell polynomials related about the $\mathcal{Y}$-constraints and $P$-conditions which turns out to be an effective tool to represent the bilinear form. On the basis of this method, the bilinear form is determined under the scaling transformation. Starting from the bilinear representation, the CKP equation is  reduced to Kaup-Kuperschmidt~equation and bidirectional Kaup-Kuperschmidt equation under the reduction of $t_3$ and $t_5$, respectively.
    \\

\end{abstract}

\maketitle

\section{Introduction}

Nonlinear evolution  equations have been widely studied due to their important applications in different areas such as nonlinear optics,  fluid mechanics, chemical physics and mathematics recently. Accordingly, many important integrable properties of soliton equations have been explored, such as soliton solutions, Lax pairs, Hamiltonian structures, symmetries, and conservation laws.
As one of the most interesting features in the nonlinear field, solitons have received more and more attention.
With the development of recent years, a variety of  approaches  have been developed to explore the soliton solution     including the inverse scattering transformation~\cite{Ablowitz1991,Hirota1974Inverse}, Painl\'eve test~\cite{Weiss1983}, Hirota's bilinear method~\cite{Hirota1971,Hirota2004,Hirota1973,Hirota1977}, binary Bell polynomials approach~\cite{Bell1934exponential,Gilson1996,Lambert2008,Lambert2001},
 B\"acklund transformations~\cite{BT1982}, Darboux transformations~\cite{Darboux1967}, $\bar{\partial}$-dressing method~\cite{zakharov1985,ablowitz1983Dbar,Dubrovsky1996,konopelchenko2013,zhu2012dressing}, Riemann-Hilbert method~\cite{deift1993}, etc. It is worth mentioning that the $\bar{\partial}$-dressing method is a fairly powerful and fundamental as well as direct method for constructing exact solution of (2+1)-dimensional integrable nonlinear evolution equations.

 The $\bar{{\partial}}$-dressing method was first proposed by Zakharov and Shabat et al.~\cite{Zakharov1974scheme}, known as a significant method, has attracted many scholars's  more attention  in mathematical physics to study solutions of (1+1) and (2+1) dimensional nonlinear evolution equations.
 The essential of this $\bar{{\partial}}$-dressing method has been explored and developed by Ablowitz, Zakharov, Fokas, and Manakov et al.~\cite{Manakov1981,ablowitz1983Dbar,zakharov1985,zakharov1988,Fokas1990dromions} and the books~\cite{konopelchenko1993,konopelchenko2013}.
  V.G.Dubrovsky et al. make a lot of contributions in this field involved rational solutions, line solutions and exact multiple pole solutions and so on~\cite{Dubrovsky1996,Dubrovskii2003construction}.  Ablowitz, Beals, Coifman and Fokas also made some historic achievements to promote the development of the $\bar{{\partial}}$-dressing method~\cite{1990DSI,1981KP,1985Multi,ablowitz1983Dbar,Fokas1990dromions}. Recently, the  $\bar{{\partial}}$-dressing method has come to people's attention again because of its powerful usefulness which possesses greater advantages in solving differential equations \cite{Fokas1992,Dubrovsky2020multiKP,Zhu2022Db,kuang2015three,Luo2020dbar,Chai2022PAMS,chai2021SK,wang2021bar,LHH2021dbar,zhao2021Dbar}.
 Starting from the Lax pair, Ablowitz et al. introduced the $\bar{{\partial}}$-dressing method to study the inverse spectral problem for a (2+1)-dimensional nonlinear equation. Later, E. V. Doktorov and S. B. Leble compiled all the dressing methods into a book~\cite{Doktorov2007dressing}.
Fan Engui and Zhu Junyi et al. also applied Dbar method to study the N-soliton solutions of (1+1)-dimensional equations with zero boundary and non-zero boundary conditions~\cite{Luo2021dbar-nzero,zhu2020dbarnzero,Luo2020dbar,Luo2021KE}. More recently, this method has also been applied to analyse long time asymptotic solutions~\cite{Michael2018,yang2019long,Fan2022AM}. In a word, not only can this method be used as an independent method to solve the equation but also play an auxiliary role in other methods.

So far, a number of equations have been successfully studied in virtue of classical $\bar{\partial}$-dressing method~\cite{Doktorov2007dressing,Fokas1992,Konopelchenko1990Inverse,Dubrovsky2020multiKP,zhu2015dbar,kuang2017zhu}, and most of them are (1+1)-dimensional equations. However, for (2+1)-dimensional equations, few works have been done with the help of inverse spectral transformation and the $\bar{\partial}$ property. In the present paper, we consider the integrable (2+1)-dimensional Kaup-Kuperschmidt equation:
\begin{eqnarray}   \label{CKP}    %eqnarry表示方程组
u_t=u_{5x}+5uu_{xxx}+\frac{25}{2} u_x u_{xx} +5u^2 u_x +5u_{xxy} -5\partial_x^{-1} u_{yy} +5uu_y +5u_x \partial_x^{-1} u_y,
\end{eqnarray}
where $\partial_x^{-1}$ is an operator inverse to $\partial_x $, $\partial_x^{-1} f=\frac{1}{2}(\int_{-\infty}^x-\int_x^{\infty})f(x^{\prime}) dx^{\prime}$.
This equation was discovered firstly in \cite{Kaup1980}, and it is also called the CKP equation. When $u(x,y,t)=u(x,t)$, Eq.(\ref{CKP}) becomes Kaup-Kupershmidt equation. To our knowledge, very little is known for the (2+1)-dimensional Kaup-Kuperschmidt equation, starting from the compatibility conditions, we bring in the eigenfunctions and Green's function of the (2+1)-dimensional Kaup-Kuperschmidt equation  firstly,
specially noting that the eigenfunctions and Green's function play an important role in receiving its solution. According to the Fourier transformation and inverse Fourier transformation, the Green's function can be confirmed. By analyzing the Green's function, we find that it didn't have jump across the real axis but it is analytic in the $z$-plane. However, the eigenfunctions were not analytic, so the non-analyticity of which prevents  from making using of the Riemann-Hilbert problem. Fortunately, the $\bar{\partial}$ problem  is found to be useful for formulating scattering equations and constructing soliton solutions.

A (3+1)-dimensional  Hirota bilinear equation is  proposed by \cite{3+1-2016Hirota},
\begin{eqnarray}   \label{3+1-Hirota}    %eqnarry表示方程组
u_{yt}-u_{xxxy}-3 ( u_x u_{y})_x  -3u_{xx} +3u_{zz}=0
\end{eqnarray}
which  admits  the similar physical meaning as the Korteweg–de Vries (KdV) equation and describes the nonlinear waves in fluid mechanics, plasma physics and weakly dispersive media. A generalized (2+1)-dimensional Hirota bilinear equation has been studied in \cite{Ma2016,Hua2019} when  $y=z$, that is
\begin{eqnarray}   \label{2+1-Hirota}    %eqnarry表示方程组
u_{yt}+c_1 [u_{xxxy}+3 ( u_x u_{y}+uu_{xy} )  +3u_{xx}  \int_{-\infty}^{x} u_y dx ]+c_2 u_{yy}=0,
\end{eqnarray}
where $c_1$ and $c_2$ are  real constants. \cite{zhao2020m} extends Eq.(\ref{2+1-Hirota}) to a generalized (2+1)-dimensional nonlinear wave equation for certain nonlinear phenomena in nonlinear optics, fluid mechanics and plasma physics,
\begin{eqnarray} \label{Wave}
u_{yt}+c_1[u_{xxxy} +3(2u_xu_y+uu_{xy}) +3u_{xx} \int_{-\infty}^x u_y dx] +c_2u_{yy}+ c_3 u_{xx} =0,
\end{eqnarray}
where $c_1$, $c_2$ and $c_3$ represent the real constants. Breather, lump, N-soliton solutions and their hybrid ones for Eq.(\ref{Wave}) have been investigated by the Hirota bilinear method~\cite{zhao2020m}. Bilinear form, Lax pair and interactions of nonlinear waves for  solutions  have been studied in \cite{zhao2021bilinear}. In a word, there is no work via the inverse spectral transformation and the $\bar{\partial}$-dressing method for Eq.(\ref{Wave}). In this paper, the inverse spectra transformation and Green's function are used to construct the $\bar{\partial}$ equation, which lays a good foundation for our further solution.

%The Lax pair of Eq.(\ref{SK}), as a starting point in constructions of exact solutions with integrable equation, can be used to construct exact solutions via  the inverse spectral transformation schemas.
%The compatibility of integrable boundary conditions allows to use all machinery of inverse spectral transformation, including different dressing methods, for construction of corresponding solutions obtained in all above schemas.

The structure of this paper is organized as follows. In Sec.II, a (2+1)-dimensional Kaup-Kuperschmidt equation is studied in detail. The concepts and formulas about the $\bar{\partial}$-dressing method will be introduced and we will derive the eigenfunctions and the Green's function of the (2+1)-dimensional Kaup-Kuperschmidt  equation.  Scattering equation of $\bar{\partial}$ form  is also discussed to construct the solution formally of Eq.(\ref{CKP}) with the help of the $\bar{\partial}$-dressing method and the symbolic computation. In Sec.III, the reduced problem of (2+1)-dimensional Kaup-Kuperschmidt equation is studied by means of the binary Bell polynomials related about the $\mathcal{Y}$-constraints and $P$-conditions. Starting from the bilinear representation, the CKP equation is  reduced to Kaup-Kuperschmidt~equation and bidirectional Kaup-Kuperschmidt equation under the reduction of $t_3$ and $t_5$, respectively. In Sec.IV, a generalized (2+1)-dimensional nonlinear wave equation for certain nonlinear phenomena in nonlinear optics, fluid mechanics and plasma physics is also investigated by the $\bar{\partial}$-dressing method in a similar way to the (2+1)-dimensional Kaup-Kuperschmidt  equation.
Based on the above sections, the conclusion will be addressed in the end.

\section{(2+1)-dimensional Kaup-Kuperschmidt equation }

The necessary notations of the $\bar{{\partial}}$-dressing method is introduced firstly to investigate its solution for (2+1)-dimensional nonlinear  equations and their related properties. If considering a matrix $\bar{{\partial}}$ problem
\begin{equation}  \label{R1}
\bar{{\partial}} \psi=\psi R,
\end{equation}
where $\psi(x,t,z)$ and $R(x,t,z)$ are $2\times2$ matrix, with a boundary condition $\psi(x,t,z)\rightarrow I$,\,$z\rightarrow \infty$, then a solution of the $\bar{{\partial}}$ problem ~(\ref{R1})~with the canonical normalization can be written as
\begin{equation}  \label{R2}
\psi(x,t,z)=I +\frac{1}{2 \pi i} \int\int\frac{\psi(\zeta)R(\zeta)}{\zeta-z}d\zeta \wedge d\bar{\zeta} .
\end{equation}
 By the way, the representation enables us to write the related properties  in the following:
\begin{eqnarray}  \label{R3}
&& \bar{{\partial}} (\psi R)=(\bar{{\partial}} \psi)R+ \psi (\bar{{\partial}}R),  \\
&& (\bar{{\partial}} \psi) _ t= \psi _t R+ \psi R_t .
\end{eqnarray}

It is known that the $\bar{\partial}$-dressing method is a very useful method to construct the solution of (2+1)-dimensional equations. The key step to get the solution by means of the $\bar{\partial}$-dressing method is shown. Once given the Lax representation of a nonlinear evoluton equation, the Jost solution is obtained, based on the result, the Green's function $G$ is found. Taking advantage of the convolution, we can get the expression of the fundamental solution which impled two linearly independent eigenfunctions of the equation. In this way, we transform the problem of exploring the solution of the equation into the problem of searching for the Green's function and the eigenfunction.
Then the Green's function $G$ is determined by making use of the residue theorem.
 In order to search for a new $\bar{\partial}$ problem, the eigenfunction and the scattering equation are discussed.
 Furthermore, a solution of the nonlinear evolution equation is received by virtue of comparing the solution of the $\bar{\partial}$ problem and eigenfunction. Finally,  the relationship between the scattering data and time is deduced by the derivative of the new $\bar{\partial}$ problem with respect to time.

\subsection{eigenfunctions and Green's function}

In this section, the (2+1)-dimensional Kaup-Kuperschmidt equation  will be considered under the inverse spectral transformation in virtue of the eigenfunctions and Green's function of its Lax pair. The Compatibility condition of Eq.(\ref{CKP}) admits the linear system
\begin{eqnarray}
&& ( \partial_{y}+\partial_{x}^{3} + u \partial_x +\frac{1}{2}u_x )\psi =0,     \label{Lax1}      \\
&& [9 \partial_x^5+ 15 u\partial_x^3 +\frac{45}{2} u_x \partial_x^2 +(\frac{35}{2}  u_{xx} +5u^2 -5 \partial_x^{-1} u_y)\partial_x +(5uu_x -\frac{5}{2}u_y +5u_{xxx} ) +\partial_t +\alpha ] \psi =0,  \label{Lax2}
\end{eqnarray}
where the index $\partial_x^n=\frac{\partial^n}{\partial x^n}, \cdots$ and~$\alpha $~ is an arbitrary constant.
Assuming that potential function~$u(x,y)$~approaches~$0$~fast enough  as $|x|+|y|\rightarrow \infty$~, in other words, the potential function ~$u(x,y)$ belongs to space $S(R)$ which is a Schwartz space
\begin{eqnarray}
S(R)=\{f\in C^{\infty}(R), \|f \|_{\gamma,\beta}=sup |x^\gamma \partial^{\beta}f(x)| < \infty ,\gamma,\beta \in Z_{+} \}  ,     \label{S}
\end{eqnarray}
then there is a Jost solution for Eq.(\ref{Lax1}) which  satisfies the asymptotic condition
\begin{eqnarray}
\psi(x,y) = e^{izx+iz^3y} +O(1),\,\, |x|,|y|\rightarrow \infty .   \label{p1}
\end{eqnarray}
 For the sake of introducing the new Lax pair with $z$ as the spectral parameter, we take the transformation
\begin{eqnarray}
\phi(x,y,z)=\psi(x,y)e^{-izx-iz^3y},\,\,    \label{p2}
\end{eqnarray}
then an new asymptotic condition of the new parameter can be expressed as
\begin{eqnarray}
\phi(x,y,z)\rightarrow 1 ,\,\, |x|,|y|\rightarrow \infty    .\label{p3}
\end{eqnarray}
Under the transformation (\ref{p2}), the Lax pair~(\ref{Lax1})-(\ref{Lax2})~can be transformed into
\begin{eqnarray}
&& \phi_{xxx}+3iz\phi_{xx}-3z^2\phi_{x}+u \phi_x +izu\phi +\frac{1}{2} u_x \phi +\phi_{y}=0 ,     \label{Lax3}      \\
&& 9\phi_{xxxxx}+45iz\phi_{xxxx}-90z^2\phi_{xxx}-90iz^3\phi_{xx}+15u\phi_{xxx}+45izu\phi_{xx}+45z^4\phi_{x}+9iz^5\phi  -45z^2u\phi_{x}-15iz^3u\phi \nonumber\\
&& \,\,\,  + \frac{45}{2} u_x (\phi_x +2iz \phi_x -z^2 \phi  ) +iz( \frac{35}{2}u_{xx}+5u^2-5\partial_{x}^{-1}u_y) \phi   + (5uu_x -\frac{5}{2}u_y +5u_{xxx})\phi + \phi_t +\alpha \phi=0 .\label{Lax4}
\end{eqnarray}
Let's consider the Green's function of Eq.(\ref{Lax3}) so that we are able to look for a bounded function $\phi(x,y,z)$,
\begin{eqnarray}
G_{xxx}+3izG_{xx}-3z^2G_{x}+G_{y}=\delta (x)\delta(y)  . \label{Lax5}
\end{eqnarray}
By making use of  the Fourier transformation of the multivariate functions and the properties of the $\delta$ function, it's easy to calculate
\begin{eqnarray}
\hat{G}(\xi,\eta,z)=\frac{1}{2\pi} \frac{1}{-i\xi^3-3iz\xi^2-3iz^2\xi+i\eta}     ,   \label{FG}
\end{eqnarray}
the Fourier inverse transformation is also utilized, we can acquire the Green's function  naturally
\begin{eqnarray}
{G}(x,y,z)=\frac{1}{4\pi^2} \int_{-\infty}^{\infty} \int_{-\infty}^{\infty}
\frac{e^{i(\xi x+\eta y)}}{-i\xi^3-3iz\xi^2-3iz^2\xi+i\eta}d\xi d\eta    .\label{FG1}
\end{eqnarray}
So the general solution of Eq.(\ref{Lax3}) is the convolution of the Green's function $G(x,y,z)$ and $-u(x,y)(\phi_x+iz\phi)-\frac{1}{2} u_x \phi$, that is
\begin{eqnarray} \label{GS}
&& \phi(x,y,z)            =G(x,y,z)*[-u(x,y)(\phi_x (x,y,z)+iz\phi(x,y,z))- \frac{1}{2} u_x \phi]              \nonumber\\
&&=\int_{-\infty}^{\infty} \! \int_{-\infty}^{\infty} \! \! G(x-x^{\prime}\! , \!y-y^{\prime},\!z)[-u(x^{\prime},y^{\prime})(\phi_{x^{\prime}} (x^{\prime},y^{\prime},z)+iz\phi(x^{\prime},y^{\prime},z)) \! -\!  \frac{1}{2} u_{x^{\prime}}( x^{\prime},y^{\prime}) \phi (x^{\prime},y^{\prime},z)  ]dx^{\prime} dy^{\prime}.
\end{eqnarray}

For the background of zero potential solution $u(x,y)=0$, direct calculation shows that there are two linearly independent eigenfunctions
\begin{eqnarray}
M_0=1, N_0=e^{-2iz_1 x-2iz_1^3 y + 6i z_1 z_2^2y } ,
\end{eqnarray}
where $z=z_1+iz_2$ is complex. Hence, the two eigenfunctions corresponding to the general potential can be expressed as
\begin{eqnarray} \label{M}
&& M(x,y,z)=1+    \nonumber\\
&& \! \int_{-\infty}^{\infty} \! \int_{-\infty}^{\infty}\!
G(x-x^{\prime} \!,\! y-y^{\prime}\!,\! z) [-u(x^{\prime},y^{\prime})(M_{x^{\prime}} (x^{\prime},y^{\prime},z)+iz M(x^{\prime},y^{\prime},z)) \! -\!  \frac{1}{2} u_{x^{\prime}}( x^{\prime},y^{\prime}) M (x^{\prime},y^{\prime},z)  ] dx^{\prime} dy^{\prime} ,
\end{eqnarray}
and
\begin{eqnarray} \label{N}
&& N(x,y,z)=e^{-2iz_1(x+(z_1^2-3z_2^2)y)}+  \nonumber\\
&&  \int_{-\infty}^{\infty} \int_{-\infty}^{\infty}
G(x-x^{\prime},y-y^{\prime},z)[-u(x^{\prime},y^{\prime})(N_{x^{\prime}} (x^{\prime},y^{\prime},z)+iz N(x^{\prime},y^{\prime},z)) \! -\!  \frac{1}{2} u_{x^{\prime}}( x^{\prime},y^{\prime}) N (x^{\prime},y^{\prime},z)  ] dx^{\prime} dy^{\prime} ,   \nonumber\\
\end{eqnarray}
where the Green's function admits
\begin{eqnarray}  \label{G}
G(x-x^{\prime},y-y^{\prime},z) =\frac{1}{4\pi^2} \int_{-\infty}^{\infty} \int_{-\infty}^{\infty}
\frac{e^{i(\xi(x-x^{\prime})+\eta (y-y^{\prime}))}}{-i\xi^3-3iz\xi^2-3iz^2\xi+i\eta}d\xi d\eta   \nonumber\\
=\frac{1}{2\pi} \int_{-\infty}^{\infty} g(\xi,y-y^{\prime},z) e^{i\xi(x-x^{\prime})}d\xi ,
\end{eqnarray}
with
\begin{eqnarray}  \label{g}
g(\xi,y-y^{\prime},z)=\frac{1}{2\pi i}\int_{-\infty}^{\infty} \frac{e^{i \eta (y-y^{\prime})}}{\eta-\xi^3-3z\xi^2-3z^2\xi} d\eta ,
\end{eqnarray}
it's obvious that the above  formula has first order pole $\eta_1=\xi (\xi^2+3z\xi+3z^2) $.

In the upper half plane, a sufficiently large semicircle  $C_R:\eta=Re^{i\theta}, R>\mid \eta_1 \mid, 0\leq \theta \leq \pi$ is made when $y-y^{\prime}>0$.  Hence, $[-R,R]\cup C_R$ forms a closed curve with the counterclockwise direction. If $\eta_1$ is in the upper half plane, then $\eta_1$ is inside the curve, otherwise, if $\eta_1$ is in the lower half plane, the integrand is analytic in the curve.  Based on the residue theorem, the following formula is established
\begin{eqnarray}  \label{Res}
\frac{1}{2\pi i} \int_{-R}^{R} \frac{e^{i\eta (y-y^{\prime})}}{\eta-\xi^3-3z\xi^2-3z^2\xi}d\eta+\frac{1}{2\pi i} \int_{C_R} \frac{e^{i\eta (y-y^{\prime})}}{\eta-\xi^3-3z\xi^2-3z^2\xi}d\eta=\underset{\eta=\eta_1} {Res}[\frac{e^{i\eta (y-y^{\prime})}}{\eta-\xi^3-3z\xi^2-3z^2\xi}] .
\end{eqnarray}
When $R\rightarrow \infty$, we can observe that the limit of  the second integral on the left-hand side is 0 by the Jordan theorem, therefore, the above formula admits
\begin{eqnarray}  \label{g1}
&&  g(\xi,y-y^{\prime},z)=\underset{\eta=\eta_1} {Res}[\frac{e^{i\eta (y-y^{\prime})}}
    {\eta-\xi^3-3z\xi^2-3z^2\xi}] \nonumber \\
&&   =
\begin{cases}
 e^{i \xi (\xi^2 +3z\xi +3z^2)(y-y^{\prime})} , 6z_1z_2 \xi +3z_2 \xi^2  > 0 , \\
0   , \quad\quad\quad \quad \quad\quad\quad\quad   6z_1z_2 \xi +3z_2 \xi^2  < 0  ,
\end{cases}      \nonumber \\
&& =H[\xi (6z_1z_2+3z_2 \xi )(y-y^{\prime})] e^{i\xi (\xi^2 +3z \xi +3z^2)(y-y^{\prime})} ,
\end{eqnarray}
where $H(\cdot)$ is the Heaviside function.
However, when $y-y^{\prime}<0$,  the following result is calculated  similarly
\begin{eqnarray}  \label{g2}
 &&  g(\xi,y-y^{\prime},z)=
\begin{cases}
 - e^{i \xi (\xi^2 +3z\xi +3z^2)(y-y^{\prime})} , 6z_1z_2 \xi +3z_2 \xi^2  < 0 , \\
0   , \quad\quad\quad \quad \quad \quad\quad\quad\quad 6z_1z_2 \xi +3z_2 \xi^2  > 0  ,
\end{cases}      \nonumber \\
&& = -H[\xi (6z_1z_2+3z_2 \xi )(y-y^{\prime})] e^{i\xi (\xi^2 +3z \xi +3z^2)(y-y^{\prime})} .
\end{eqnarray}
Combining the two formulas (\ref{g1}) and (\ref{g2}) yields,
\begin{eqnarray}  \label{g3}
  g(\xi,y-y^{\prime},z)= sgn(y-y^{\prime})e^{i\xi (\xi^2+3z\xi+3z^2)(y-y^{\prime})} ,
\end{eqnarray}
Thus, the Green's function holds
\begin{eqnarray}  \label{Gg}
G(x-x^{\prime},y-y^{\prime},z) =\frac{1}{2\pi} \int_{-\infty}^{\infty} sgn(y-y^{\prime})    H[\xi (6z_1z_2+3z_2 \xi )(y-y^{\prime})]     e^{i\xi (\xi^2+3z\xi+3z^2)(y-y^{\prime})+i\xi(x-x^{\prime})}d\xi .
\end{eqnarray}

\subsection{scattering equation and $\bar{\partial}$ problem }
Because the Green's function has no jump across the real axis and this function is analytic in the $z$-plane. However, the eigenfunctions $M,N$ are not analytic in the same plane, which prevents us using the Riemann-Hilbert method. That is why  the $\bar{\partial}$-dressing method is applied to study solutions of the (2+1)-dimensional Kaup-Kuperschmidt equation.
Ablowitz et al. also found that the $\bar{\partial}$ problem could be employed to formulate scattering equations and to solve the inverse scattering problem~\cite{Ablowitz1991,ablowitz1983Dbar}.
Hence, we calculate the $\bar{\partial}$ derivative firstly as follows:
\begin{eqnarray}  \label{dbarM}
&& \bar{\partial}M(x,y,z)=  \nonumber    \\
&& \int_{-\infty}^{\infty} \int_{-\infty}^{\infty}[\bar{\partial}G(x-x^{\prime},y-y^{\prime},z)][-u(x^{\prime},y^{\prime})(M_{x^{\prime}} (x^{\prime},y^{\prime},z)+izM(x^{\prime},y^{\prime},z))-\frac{1}{2} u_{x^{\prime}}M(x^{\prime},y^{\prime},z ) ]dx^{\prime} dy^{\prime}          \nonumber    \\
&& +\int_{-\infty}^{\infty} \int_{-\infty}^{\infty}G(x-x^{\prime},y-y^{\prime},z)[-u(x^{\prime},y^{\prime})(\bar{\partial}(M_{x^{\prime}} (x^{\prime},y^{\prime},z)+izM(x^{\prime},y^{\prime},z))\!-\! \frac{1}{2} u_{x^{\prime}} \bar{\partial}  M(x^{\prime},y^{\prime},z )   ]dx^{\prime} dy^{\prime}\! ,
\end{eqnarray}
with
\begin{eqnarray}  \label{dbarG}
&& \bar{\partial}G(x-x^{\prime},y-y^{\prime},z)=\frac{1}{4\pi^2} \int_{-\infty}^{\infty} \int_{-\infty}^{\infty}
e^{i(\xi(x-x^{\prime})+\eta (y-y^{\prime}))}\bar{\partial}{\frac{1}{ i\eta-i\xi^3-3iz\xi^2-3iz^2\xi}}d\xi d\eta  ,  \nonumber\\
&&\quad\quad \quad \quad\quad \quad =\frac{1}{4\pi} \int_{-\infty}^{\infty} \int_{-\infty}^{\infty}  \frac{e^{i(\xi(x-x^{\prime})+\eta (y-y^{\prime}))}}{-3i\xi}
\delta(z_1^2-z_2^2+z_1\xi+\frac{\xi^2}{3}-\frac{\eta}{3\xi})\delta(2z_1z_2+z_2\xi)    d\xi d\eta ,  \nonumber\\
&&\quad\quad \quad \quad\quad \quad =\frac{i}{4\pi\mid z_2 \mid} sgn(-z_1) e^{i(\xi(x-x^{\prime})+\eta(y-y^{\prime}))}\mid_{{\xi=-2z_1};\eta
     =-2z_1^3+6z_1z_1^2}   ,    \nonumber\\
&&\quad\quad \quad\quad\quad \quad  =\frac{i}{4\pi\mid z_2 \mid} sgn(-z_1) e^{-2iz_1[(x-x^{\prime})+(z_1^2-3z_2^2)(y-y^{\prime})]}       .
\end{eqnarray}
It leads to the following equation directly
\begin{eqnarray}  \label{dbarM1}
&& \bar{\partial}M(x,y,z)=F(z_1,z_2)  e^{-2iz_1[x+(z_1^2-3z_2^2)y]}  \nonumber\\
&&  +\int_{-\infty}^{\infty} \int_{-\infty}^{\infty}G(x-x^{\prime},y-y^{\prime},z)[-u(x^{\prime},y^{\prime})\bar{\partial}(M_{x^{\prime}} (x^{\prime},y^{\prime},z)+izM(x^{\prime},y^{\prime},z))-\! \frac{1}{2} u_{x^{\prime}} \bar{\partial}  M(x^{\prime},y^{\prime},z )   ]dx^{\prime} dy^{\prime},
\end{eqnarray}
where the function $F(z_1,z_2)$ reads
\begin{eqnarray}  \label{F}
&& F(z_1,z_2)=\frac{i}{4\pi\mid z_2 \mid} sgn(-z_1) \int_{-\infty}^{\infty} \int_{-\infty}^{\infty}[-u(x^{\prime},y^{\prime})(M_{x^{\prime}} (x^{\prime},y^{\prime},z)+izM(x^{\prime},y^{\prime},z))   \nonumber\\
&&  -\frac{1}{2} u_{x^{\prime}}M(x^{\prime},y^{\prime},z )     ]
 e^{2iz_1(x^{\prime}+(z_1^2-3z_2^2)y^{\prime})}dx^{\prime} dy^{\prime}   .
\end{eqnarray}

Let's multiply both sides of Eq.(\ref{N}) by $F(z_1,z_2)$  and subtract Eq.(\ref{dbarM1}), one can receive the  scattering equation in the form of the linear $\bar{\partial}$ problem by  comparing Eq.(\ref{dbarM1}) with the known Eq.(\ref{N}), that is
\begin{eqnarray}  \label{FN-dbarM}
\bar{\partial}M(x,y,z)= F(z_1,z_2)N(x,y,z),
\end{eqnarray}
and
\begin{eqnarray}  \label{dbarM2}
\bar{\partial}[M_{x} (x,y,z)+izM(x,y,z)]=F(z_1,z_2)(N_x (x,y,z)+izN(x,y,z))  .
\end{eqnarray}
A symmetry (closure) relation of Green's function is concentrated on to express the eigenfunction $N$ in term of $M$, the symmetry relation is taken as
\begin{eqnarray}  \label{Gz}
&& G(x-x^{\prime},y-y^{\prime},-\bar{z}) =\frac{1}{4\pi^2} \int_{-\infty}^{\infty} \int_{-\infty}^{\infty}
\frac{e^{i(\xi(x-x^{\prime})+\eta (y-y^{\prime}))}}{-i\xi^3+3i\bar{z}\xi^2-3i\bar{z}^2\xi+i\eta}d\xi d\eta   \nonumber\\
&& =\frac{1}{4\pi^2} \int_{-\infty}^{\infty} \int_{-\infty}^{\infty}
\frac{e^{i(\xi(x-x^{\prime})+\eta (y-y^{\prime}))}}{-i(\xi-2z_1)^3-3iz(\xi-2z_1)^2-3iz^2(\xi-2z_1)+i(\eta-2z_1^3+6z_1z_2^2)}d\xi d\eta ,
\end{eqnarray}
where $z=z_1+iz_2$. By taking the transformation $\xi-2z_1 \rightarrow \xi$, $\eta-2z_1^3+6z_1z_2^2 \rightarrow \eta$, the symmetry equality of the Green's function is denoted as
\begin{eqnarray}  \label{Gz}
G(x-x^{\prime},y-y^{\prime},-\bar{z}) =G(x-x^{\prime},y-y^{\prime},z) e^{2iz_1[(x-x^{\prime})+(z_1^2-3z_2^2)(y-y^{\prime})]}  .
\end{eqnarray}
Replacing $z$ with $-\bar{z}$ in Eq.(\ref{M}) and comparing with the above formula, yields that
\begin{eqnarray}  \label{M-z}
&& M(x,y,-\bar{z}) e^{-2iz_1[x+(z_1^2-3z_2^2)y]}=e^{-2iz_1[x+(z_1^2-3z_2^2)y]}     \nonumber\\
&& +\int_{-\infty}^{\infty} \int_{-\infty}^{\infty}
 G(x-x^{\prime},y-y^{\prime},z)[-u(x^{\prime},y^{\prime})(M_{x^{\prime}} (x^{\prime},y^{\prime},-\bar{z})-i \bar{z} M(x^{\prime},y^{\prime},-\bar{z}))\nonumber\\
&& - \frac{1}{2} u_{x^{\prime}} M(x^{\prime},y^{\prime},-\bar{z}) ] e^{-2iz_1[x^{\prime}+(z_1^2-3z_2^2)y^{\prime}]}   dx^{\prime} dy^{\prime},
\end{eqnarray}
by observing the eigenfunction $N$, the other relation can be described as
\begin{eqnarray}  \label{MN}
N(x,y,z)=M(x,y,-\bar{z}) e^{-2iz_1[x+(z_1^2-3z_2^2)y]},
\end{eqnarray}
that is, the $\bar{\partial}$ problem for the eigenfunctions $M$ and $N$ is further expressed in terms of the scattering data:
\begin{eqnarray}  \label{MNdbar}
\bar{\partial}M(x,y,z)=F(z_1,z_2)M(x,y,-\bar{z}) e^{-2iz_1[x+(z_1^2-3z_2^2)y]} .
\end{eqnarray}
%It should be stressed that the above calculation is valid under the assumption that there are no nontrivial solutions of the homogeneous integral equation obtained from Eq.(\ref{N}).

\subsection{Inverse spectral problem}

It is well known that the inverse spectral problem can be solved by using the Cauchy-Green formula which is shown below
\begin{eqnarray}  \label{Msol}
M(x,y,z)=1+\frac{1}{2\pi i} \int \int \frac{d\zeta \wedge d\bar{\zeta}}{\zeta-z}
F(\zeta_1,\zeta_2)M(x,y,-\bar{\zeta}) e^{-2i\zeta_1[x+(\zeta_1^2-3\zeta_2^2)y]} ,
\end{eqnarray}
%By the way, we can write the solution
%\begin{eqnarray}  \label{Mxsol}
%M(x,y,z)+izM(x,y,z)=1+\frac{1}{2\pi i} \int \int \frac{d\zeta \wedge d\bar{\zeta}}{\zeta-z}
%F(\zeta_1,\zeta_2)[M_x(x,y,\zeta)+izM(x,y,\zeta)] e^{-2i\zeta_1[x+(\zeta_1^2-3\zeta_2^2)y]}
%\end{eqnarray}
where $\zeta=\zeta_1+i \zeta_2$.
In order to reconstruct the potential function $u(x,y)$, it is necessary to compare the two expressions of $M-1$,
\begin{eqnarray}  \label{doubleM-1}
&&  M(x,y,z)-1=  \nonumber\\
&& \begin{cases}
\int_{-\infty}^{\infty} \int_{-\infty}^{\infty}
G(x-x^{\prime},y-y^{\prime},z) [-u(x^{\prime},y^{\prime})(M_{x^{\prime}} (x^{\prime},y^{\prime},z)+ i z M(x^{\prime},y^{\prime},z ))
- \frac{1}{2} u_{x^{\prime}} M(x^{\prime},y^{\prime},z) ]
dx^{\prime} dy^{\prime},     \\
\frac{1}{2\pi i}\int_{-\infty}^{\infty} \int_{-\infty}^{\infty}\frac{d\zeta\wedge d\bar{\zeta}} {\zeta-z} F(\zeta_1,\zeta_2)
M(x,y,-\bar{\zeta} )e^{-2i\zeta_1(x +({\zeta_1^2-3 \zeta_2^2}y)}.
\end{cases}
\end{eqnarray}

Resort to the Green's function for $\mid z \mid\rightarrow \infty$, there is an asymptotic expression
\begin{eqnarray}  \label{Go}
&& G(x,y,z,\bar{z}) =\frac{1}{4\pi^2} \int_{-\infty}^{\infty} \int_{-\infty}^{\infty}
   \frac{e^{i(\xi x+\eta y)}}{-i\xi^3-3iz\xi^2-3iz^2\xi+i\eta}d\xi d\eta   \nonumber\\
&&    =\frac{1}{-12i\pi ^2 z^2} \int_{-\infty}^{\infty} e^{i\eta y}d\eta \int_{-\infty}^{\infty} \frac{1}{\xi}e^{i\xi x}d\xi+O(z^{-3})        \nonumber\\
&&  = -\frac{1}{6z^2}sgn(x)\delta(y) +O(z^{-3}) .
\end{eqnarray}
Besides, noting that $M(x,y,z)=1+O(z^{-1})$ and substituting Eq.(\ref{Go}) into the first one of Eq.(\ref{doubleM-1}), the asymptotic expansion can be derived as
\begin{eqnarray}  \label{M-1}
\partial_x[M(x,y,z)-1] = (\frac{i}{3z}+ \frac{4i}{3z^2})u +\frac{1}{6z^2}u_x +O(z^{-3}),
\end{eqnarray}
the similar asymptotic expression from the $\bar{\partial}$ representation is
\begin{eqnarray}  \label{M-12}
M(x,y,z)-1 =  \frac{1}{2\pi i}\int\int  (-\frac{1}{z}-\frac{\zeta}{z^2})  F(\zeta_1,\zeta_2)M(x,y,-\bar{\zeta})e^{-2i\zeta_1[x+(\zeta_1^2-3\zeta_2^2)y]}  d\zeta \wedge d\bar{\zeta}   +O(z^{-3}) .
\end{eqnarray}
By comparing Eq.(\ref{M-1}) and Eq.(\ref{M-12}), the reconstruction formula is acquired
\begin{eqnarray}  \label{u}
&& u(x,y) =  \nonumber\\
 && e^{(-2iz-8i)x} [\int \frac{3z^2i}{\pi} \partial_x \int\int (\frac{1}{z}+ \frac{\zeta}{z^2}) F(\zeta_1,\zeta_2) M(x,y,-\bar{\zeta}) e^{-2i\zeta_1(x+(\zeta_1^2-3\zeta_2^2)y)} d\zeta \wedge d\bar{\zeta}\cdot e^{(2iz+8i)x} dx +\tilde{c}],
\end{eqnarray}
where  $\tilde{c}$ is an arbitrary parameter.

Once the time evolution of spectral data is determined, a solution of the CKP hierarchy is attained. Hence, substituting Eq.(\ref{p2}) into $\bar{\partial}$-equation gives
\begin{eqnarray}  \label{psidbar}
\bar{\partial}\psi(x,y,t,z)=F(z_1,z_2,t)\psi(x,y,t,-\bar{z}) ,
\end{eqnarray}
taking the derivative of both sides of this equation with respect to $t$ yields
\begin{eqnarray}  \label{psit}
[\bar{\partial}\psi(z)]_t=F_t(z_1,z_2,t)\psi(-\bar{z})+F(z_1,z_2,t)\psi_t(-\bar{z}) .
\end{eqnarray}

While noting Eq.(\ref{Lax2}) which $\psi(z)$ is satisfied and $u\in S(R)$,  $\alpha=-9iz^5$ can be observed from the following expression
\begin{eqnarray}  \label{ alpha }
9(iz)^5 e^{izx+iz^3y} +\alpha e^{izx+iz^3y}=0 ,
\end{eqnarray}
then  it lead to the following result in touch with Eq.(\ref{psit})
\begin{eqnarray}  \label{F1}
F(z_1,z_2,t)=F(z_1,z_2)e^{(9iz^5+9i\bar{z}^5)t} ,
\end{eqnarray}
where $F(z_1,z_2,0)=F(z_1,z_2)$.
Finally, the solution (\ref{u})  can be represented as
\begin{eqnarray}  \label{ut}
&& u(x,y,t) =  \nonumber\\
 && e^{(-2iz-8i)x} [\int \frac{3z^2i}{\pi}\partial_x \int\int (\frac{1}{z}+\frac{\zeta}{z^2}) F(\zeta_1,\zeta_2) M(x,y,t,-\bar{\zeta}) e^{-2i\zeta_1(x+(\zeta_1^2-3\zeta_2^2)y)} d\zeta \wedge d\bar{\zeta}   \nonumber\\
 && \cdot e^{(2iz+8i)x+9i(z^5+ \bar{z}^5)t} dx +\tilde{c}],
\end{eqnarray}
and the formula $F(z_1,z_2)$ is shown in Eq.(\ref{F}).

\section{CKP equation and its reduction}
The well-known KP equation admits
\begin{eqnarray}
(4u_{t_3}-u_{3x} -6uu_x)_x -3u_{2t_2}=0,
\end{eqnarray}
where $t_n$ is the variable whose weight is $n$. Sometimes this equation is called the KP equation of type A (AKP equation), and the reduced problems of weight 2 and 3  have been studied~\cite{Lambert2008},  the reduced equations are given as KdV equation
\begin{eqnarray}
u_{t} -6uu_x +3u_{xxx}=0,
\end{eqnarray}
and  Boussinesq equation
\begin{eqnarray}
3u_{tt}+(u_{xx} +3u^2)_{xx}=0.
\end{eqnarray}
The reduction of weight 3 and 5 for BKP equation has  been considered as well, this two equations are Sawada Kotera equation
\begin{eqnarray}
u_{5x} +15u_xu_{xx} +15uu_{xxx} +45u^2u_x  +9u_{t}=0,
\end{eqnarray}
and Ramani equation
\begin{eqnarray}
u_{6x}+15u_{2x}u_{3x}+15u_xu_{4x}+45u_x^2u_{2x} -5(u_{3x,t}+3u_{2x}u_t+3u_x u_{xt})-5u_{2t}=0.
\end{eqnarray}
The (2+1)-dimensional Sawada Kotera equation has been studied by means of the $\bar{\partial}$-dressing method~\cite{chai2021SK}. The coupled Ramani equation has been considered by making use of binary bell polynomial method \cite{2020Ramani}.

A natural problem will be noted to explore the reduced problems of the CKP hierarchy which is also called (2+1)-dimensional Kaup-Kuperschmidt equation
\begin{eqnarray} \label{eqnCKP}
 u_t+\frac{1}{9}u_{5x}+\frac{25}{6}u_xu_{2x}+\frac{5}{3}uu_{xxx}+5u^2u_x-\frac{5}{9}\partial_x^{-1}u_{yy}  -\frac{5}{9} u_{xxy}-\frac{5}{3}u_{x}\partial_x^{-1}u_y-\frac{5}{3}uu_{y}=0.
\end{eqnarray}
The reduction of the above equation will be  studied from the bilinear form in the present paper. In order to detect its existence of linearizable representation, the main tool used is a class of generalized multi-dimensional binary Bell polynomials. To make our presentation easier to understand and more self-contained, we first fix some necessary notations on the Bell polynomials, the details refer, for instance, to Bell, Lambert and Gilson’s work~\cite{Bell1934exponential,Gilson1996,lambert2001classical,Lambert2008}.
During the early 1930s, E. T. Bell introduced three kinds of exponential polynomials, one of which is defined by
\begin{eqnarray}\label{TP}
Y_{nx}(y)\equiv Y_n(y_{x},\dots,y_{nx})=e^{-y}\partial_x^ne^y,\, \,\quad n=1,2,\dots ,
\end{eqnarray}
where $y$ is a $C^\infty$ function of $x$ and $y_{rx}=\partial_x^ry$ for $r=1,2,\dots$.

Later on, Lambert et al. generalized  one-dimensional Bell polynomials to multi-dimensional ones \cite{Gilson1996}. Let $y=y(x_1,x_2,\dots,x_n)$ be a  $C^{\infty} $ function of multi-variables, the multi-dimensional Bell polynomials are defined by
\begin{equation}
Y_{n_1x_1,n_2x_2,\dots,n_lx_l}(y)\equiv Y_{n_1,n_2,\dots,n_l}(y_{r_1x_1,\dots,r_lx_l})=e^{-y}\partial_{x_1}^{n_1}\cdots \partial_{x_l}^{n_l}e^y,
\end{equation}
where
\begin{equation}
y_{r_1x_1,\dots,r_lx_l}=\partial_{x_1}^{r_1}\cdots \partial_{x_l}^{r_l}y \,\,(r_k=0,\dots,n_k,\, k=1,\dots, l).
\end{equation}
For the special case $y=y(x,t)$,  the two-dimensional Bell polynomials yields,
\begin{equation}
Y_{x,t}=y_{x,t}+y_xy_t,\,\,Y_{2x,t}=y_{2x,t}+y_x^2y_t+y_{2x}y_t+2y_{x,t}y_x,\,\,...\, .
\end{equation}
Based on the multi-dimensional Bell polynomials, the multi-dimensional binary Bell polynomials ($\mathcal{Y}$-polynomials) are introduced as follows
\begin{equation}
\mathcal{Y}_{n_1x_1,\dots,n_lx_l}(v,w)=Y_{n_1x_1,\dots,n_lx_l}(y)\left | \right.{_{y_{r_1x_1,\dots,r_lx_l}=
\left\{
\begin{array}{l}
w_{r_1x_1,\cdots,r_lx_l},\, {\rm if}  \, r_1+\cdots +r_l\,{\rm is  \,even} ,\\
v_{r_1x_1,\cdots,r_lx_l},\,  {\rm if}  \, r_1+\cdots +r_l\,{\rm is  \, odd}.
\end{array}  \right.}}
\end{equation}
The first few formulas can be shown as
\begin{eqnarray*}  \nonumber
&&\mathcal{Y}_{x}(v)=v_x,\,  \mathcal{Y}_{x,t}(v,w)=w_{x,t}+v_x v_t,\\ \nonumber
&&\mathcal{Y}_{2x}(v,w)=w_{2x}+v_x^2,\, \mathcal{Y}_{3x}(v,w)=v_{3x}+3w_{2x}v_x+v_x^3,  \\ \nonumber
&&\mathcal{Y}_{2x,t}(v,w)=w_{2x}v_t +v_x^2 v_t +2v_x w_{x,t}+v_t w_{2x},\\ \nonumber
&&\mathcal{Y}_{4x}(v,w)=w_{4x}+4v_{3x}v_x+3w_{2x}^2+6w_{2x}v_x^2+v_x^4,\\
&&\mathcal{Y}_{5x}(v,w)=v_{5x}+10v_{3x}v_x^2+10w_{2x}v_x^3+15w_{2x}^2v_x+5w_{4x}v_x+10v_{3x}w_{2x}+v_x^5.
\end{eqnarray*}
In particular, by setting $v=0$ and $w=q$ in the binary Bell polynomials, we have the so-called $P$-polynomials
\begin{equation}
P_{n_1x_1,\dots,n_lx_l}(q)=\mathcal{Y}_{n_1x_1,\dots,n_lx_l}(v=0,w=q).
\end{equation}

On the one hand, the link between binary Bell polynomials and Hirota's bilinear expressions are given by the identity
\begin{equation}\label{BB}
(FG)^{-1}D_{x_1}^{n_1}\cdots D_{x_l}^{n_l}F\cdot G=\mathcal{Y}_{n_1x_1,\cdots,n_lx_l}(v=\ln F/G,\,w=\ln FG) ,
 \end{equation}
where operators $D_{x_1}^{n_1},\cdots ,D_{x_l}^{n_l}$ are the classical Hirota's bilinear operator defined  by
\begin{equation}
D_{x_1}^{n_1}\cdots D_{x_l}^{n_l}F\cdot G=(\partial_{x_1}-\partial_{{x_1}^{\prime}})^{n_1}\cdots(\partial_{x_l}-\partial_{{x_l}^\prime})^{n_l}F(x_1,\cdots,x_l)
G(x_1^{\prime},...,x_l^{\prime})\mid_{x_i^{\prime}=x_i}.\,\,
\end{equation}
In the particular case of $F=G$, the Eq.(\ref{BB}) can be reduced to
\begin{equation}\label{PB}
G^{-2}D_{x_1}^{n_1}\cdots D_{x_l}^{n_l}G\cdot G=\mathcal{Y}_{n_1x_1,\cdots,n_lx_l}(0,\,q=2\ln G)=P_{n_1x_1,\cdots,n_lx_l}(q).
 \end{equation}
On the other hand, the binary Bell polynomials can be decomposed into $Y$-polynomials and $P$-polynomials as
\begin{eqnarray}\label{BBD}
&&(FG)^{-1}D_{x_1}^{n_1}\cdots D_{x_l}^{n_l}F\cdot G|_{v=\ln F/G, w=\ln FG}\nonumber\\
%=\mathcal{Y}_{n_1x_1,\cdots,n_lx_l}(v,w)|_{v=\ln F/G, w=\ln FG}\nonumber \\
&&=\mathcal{Y}_{n_1x_1,\cdots,n_lx_l}(v, v+q)|_{q=2\ln G}\nonumber \\
&&=\sum_{p_{1}=0}^{n_{1}}\cdots \sum_{p_{l}=0}^{n_{l}} \left(
                                              \begin{array}{c}
                                                n_{1} \\
                                                p_{1} \\
                                              \end{array}
                                            \right)\cdots \left(
                                              \begin{array}{c}
                                                n_{l} \\
                                                p_{l} \\
                                              \end{array}
                                            \right)
                                            Y_{(n_{1}-p_{1})x_{1},\dots,(n_{l}-p_{l})x_{l}}(v)P_{n_1x_1,\dots,n_lx_l}(q).
\end{eqnarray}

\begin{thm} \label{thm} \quad  Under the transformation $u=\frac{1}{2}({\rm ln}f)_{xx}$, the (2+1)-dimensional Kaup-Kuperschmidt equation  can be bilinearized into
\begin{eqnarray} \label{CKP-B}
&&(D_{x}^{6}+144D_{x}D_{t}-80D_{y}^{2}-20D_{x}^3D_{y})f\cdot f-90D_x^2f\cdot g=0  , \nonumber\\
&&(D_{x}^4-4D_{x}D_{y})f\cdot f +6fg=0  ,
\end{eqnarray}
where $f=f(x,t,z),g=g(x,t,z)$, and $D$ is Hirota's bilinear operator.
\end{thm}

{\bf Proof} \quad The invariance of Eq.\eqref{eqnCKP} under the scale transformation
\begin{eqnarray}
x \rightarrow \lambda x,\,\,t \rightarrow\lambda^5 t,\,\,y\rightarrow\lambda^3 y,\,\,u\rightarrow \lambda^{-2} u,
\end{eqnarray}
shows that $u$ can be related to a dimensionless field $q=q(x,t,z)$ by setting
\begin{equation}\label{DVT}
u=cq_{xx},
\end{equation}
where $q=2lnf$ and $c$ is a dimensionless free constant to be determined. Then Eq.\eqref{eqnCKP} becomes
\begin{eqnarray} \label{1}
 E(q)\equiv 9q_{xxt}+q_{7x}+\frac{75}{2}cq_{3x}q_{4x}+15cq_{2x}q_{5x}+45c^2q_{2x}^2q_{3x}
 -5q_{xyy}-5q_{4x,y}-15cq_{3x}q_{xy}-15cq_{2x}q_{2x,y}=0.
\end{eqnarray}
Compared with $P$-polynomials, which implies that we should require  $c=\frac{1}{2}$, the result equation can be cast into a combination form
\begin{eqnarray}   \label{1-1}
 E(q)&& \equiv 9q_{xxt}+q_{7x}+\frac{75}{4}q_{3x}q_{4x}+\frac{15}{2}q_{2x}q_{5x}++\frac{45}{4}q_{2x}^2q_{3x}    \nonumber\\
 &&\quad\, -5q_{xyy}-5q_{4x,y}-\frac{15}{2}q_{3x}q_{xy}-\frac{15}{2}q_{2x}q_{2x,y}=0    \nonumber\\
 &&\equiv  9q_{xxt}+\partial_x(\frac{1}{16}P_{6x}+\frac{15}{16}(\partial_x^2+q_{2x})P_{4x})         \nonumber\\
 &&\quad\, -5q_{xyy}-5q_{4x,y}-\frac{15}{2}q_{3x}q_{xy}-\frac{15}{2}q_{2x}q_{2x,y}     \nonumber\\
 && \equiv \partial_x(9q_{xt})+\partial_x(\frac{1}{16}P_{6x}+\frac{15}{16}(\partial_x^2+q_{2x})P_{4x})         \nonumber\\
 &&\quad\, -\partial_x(5q_{yy}) -\partial_x(5P_{3x,y})+\partial_x(\frac{15}{2}q_{2x}q_{xy})=0.
\end{eqnarray}
In terms of auxiliary function
\begin{eqnarray} \label{2}
P_{4x}=4P_{xy}-6\frac{g}{f},
\end{eqnarray}
and the link between binary Bell polynomials and Hirota's bilinear operators, it holds
\begin{eqnarray} \label{3}
D_{x}^4f\cdot f=4D_{x}D_yf\cdot f-6fg .
\end{eqnarray}
In Eq.(\ref{1-1}), because of the identity
\begin{eqnarray} \label{3-1}
&& (\partial_x^2+q_{2x})P_{4x}=(\partial_x^2+q_{2x})(q_{4x}+3q_{2x}^2)=q_{6x}+6q_{3x}^2+7q_{2x}q_{4x}+3q_{2x}^3  , \nonumber\\
&& \frac{1}{16}P_{6x}=\frac{1}{16}q_{6x}+\frac{15}{16}q_{2x}q_{4x}+\frac{15}{16}q_{2x}^3 ,
\end{eqnarray}
the following equation is derived
\begin{eqnarray} \label{3-3}
\partial_x(\frac{15}{16}(\partial_x^2+q_{2x})P_{4x}+\frac{1}{16}P_{6x})=
q_{7x}+\frac{75}{4}q_{3x}q_{4x}+\frac{15}{2}q_{2x}q_{5x}+\frac{45}{4}q_{2x}^2q_{3x}  .
\end{eqnarray}
Thus, by integrating both sides of Eq.(\ref{1-1}), the new equation can be rewritten as
\begin{eqnarray} \label{4}
 E(q)&&= 9P_{xt}+\frac{1}{16}P_{6x}+\frac{15}{16}(\partial_x^2+q_{2x})P_{4x}-5P_{yy}-5P_{3x,y}+\frac{15}{2}q_{2x}q_{xy}            \nonumber\\
 &&= 9P_{xt}+\frac{1}{16}P_{6x}-5P_{3x,y}+\frac{15}{16}[(\partial_x^2+q_{2x})\cdot 4q_{xy}-6(\partial_x^2+q_{2x})\frac{g}{f}]      -5P_{yy}+\frac{15}{2}q_{2x}q_{xy}=0  .
\end{eqnarray}
Finally, we arrive at
\begin{eqnarray} \label{5}
 E(q)= 9P_{xt}+\frac{1}{16}P_{6x}-5P_{yy}-\frac{5}{4}P_{3x,y}-\frac{45}{8}\frac{g}{f} \mathscr{Y}_{2x}=0,
\end{eqnarray}
noticing the relationship  between binary Bell polynomials and Hirota's bilinear expressions  as well as multiplying both sides  by $f^2$ and 16,  it is clear   that  Eq.\eqref{eqnCKP} have the bilinear representation   Eq.\eqref{CKP-B} from  Eq.\eqref{5} and  Eq.\eqref{2}.
$\square$

Starting from the bilinear representation, the reduction is discussed. It is easy to find that the weight of $y$ is 3 in Eq.\eqref{CKP-B}, for the sake of convenience, it is taken as $t_3$, the same weight for $t$ is 5 and it is taken as $t_5$. Hence, the bilinear form can be rewritten as
\begin{eqnarray} \label{CKP-D}
&&(D_{x}^{6}+144D_{x}D_{t_5}-80D_{t_3}^{2}-20D_{x}^3D_{t_3})f\cdot f-90D_x^2f\cdot g=0, \nonumber\\
&&(D_{x}^4-4D_{x}D_{t_3})f\cdot f +6fg=0.
\end{eqnarray}
On the one hand, the $t_3$ reduction  is performed as
\begin{eqnarray} \label{CKP-Dt3}
&&(D_{x}^{6}+144D_{x}D_{t_5})f\cdot f-90D_x^2f\cdot g=0 ,\nonumber\\
&&D_{x}^4f\cdot f +6fg=0,
\end{eqnarray}
by further making use of  identity  Eq.\eqref{PB} and \begin{eqnarray} \label{6}
(\partial_x^2+q_{2x})\frac{g}{f}=(\partial_x^2+q_{2x})e^v=(v_x+v_{xx}+q_{2x})e^v=\frac{g}{f} \mathscr{Y}_{2x} ,
\end{eqnarray}
where $v={\rm ln}{g}/{f}$~, and according to $P_{4x}=-6g/f$, which demonstrates that
\begin{eqnarray} \label{7}
16q_{6x}+120q_{2x}q_{4x}+60q_{2x}^3+144q_{xt}+90q_{3x}^2=0  .
\end{eqnarray}
To proceed, taking the derivative of both sides of the above equation, it is easy to deduce that
\begin{eqnarray} \label{8}
16q_{7x}+300q_{3x}q_{4x}+120q_{2x}q_{5x}+180q_{2x}^2q_{3x}+144q_{xxt}=0  ,
\end{eqnarray}
which implies that
\begin{eqnarray} \label{9}
\frac{1}{9}w_{5x}+\frac{25}{6}w_{x}w_{2x}+\frac{5}{3}ww_{3x}+5w^2w_{x}+w_{t}=0 ,
\end{eqnarray}
on account of $w=\frac{1}{2}q_{xx}$ and $q=2{\rm ln}f$.
Therefore, Eq.\eqref{9} is the~Kaup-Kuperschmidt~equation or KK equation.

On the other hand, the $t_5$ reduction  possesses
\begin{eqnarray} \label{CKP-D}
&&(D_{x}^{6}-80D_{t_3}^{2}-20D_{x}^3D_{t_3})f\cdot f-90D_x^2f\cdot g=0 , \nonumber\\
&&(D_{x}^4-4D_{x}D_{t_3})f\cdot f +6fg=0  .
\end{eqnarray}
On account of  identity  Eq.\eqref{PB} and
\begin{eqnarray} \label{10}
&&D_x^2f\cdot g=f^{-1}g\cdot \mathscr{Y}_{2x}=(v_x+v_{xx}+q_{2x})\frac{g}{f}=(\partial_x^2+q_{2x}) \frac{g}{f}  \nonumber\\
&&\quad\,\,\quad \quad =(\partial_x^2+q_{2x})\frac{1}{6}(4P_{x,t_3}-P_{4x}),
\end{eqnarray}
it follows from Eq.\eqref{CKP-D} that
\begin{eqnarray}
P_{6x}-80P_{2t_3}-20P_{3x,t_3}-90\frac{g}{f}\cdot \mathscr{Y}_{2x}=0,
\end{eqnarray}
that is
\begin{eqnarray} \label{11}
16q_{6x}+120q_{2x}q_{4x}+60q_{2x}^3-80q_{2t_3}-80q_{3x,t_3}-120q_{2x}q_{x,t_3}+90q_{3x}^2=0.
\end{eqnarray}
Furthermore, taking the derivative about the above equation yields
\begin{eqnarray} \label{12}
&& 16q_{7x}+120q_{3x}q_{4x}+120q_{2x}q_{5x}+180q_{2x}^2q_{3x}-80q_{x,2t_3}-80q_{4x,t_3}-120q_{3x}q_{x,t_3}  \nonumber\\
&& -120q_{2x}q_{xx,t_3}+180q_{3x}q_{4x}=0 ,
\end{eqnarray}
which describes the bidirectional Kaup-Kuperschmidt equation or bKK equation
\begin{eqnarray} \label{13}
 \frac{1}{9}u_{6x}+\frac{25}{6}u_{2x}u_{3x}+\frac{5}{3}u_{x}u_{4x}+5u_x^2u_{2x}-\frac{5}{9}u_{2t_3}-\frac{5}{9}u_{3x,t_3}
 -\frac{5}{3}u_{2x}u_{t_3}-\frac{5}{3}u_{x}u_{x,t_3}=0 , \nonumber\\
\end{eqnarray}
where $u=\frac{1}{2}q_{x}$ and $q=2{\rm ln}f$ have been taken into consideration.

\section{Generalized (2+1)-dimensional nonlinear wave equation}

\subsection{eigenfunctions and Green's function}

A generalized (2+1)-dimensional nonlinear wave equation  in nonlinear optics, fluid mechanics, plasma physics is Eq.(\ref{Wave}).
In order to acquire the inverse spectral transformation by the eigenfunctions and Green's function,  starting from the Lax pair of Eq.(\ref{Wave}) which possesses the linear system
\begin{eqnarray} \label{LaxWa}
&& (3c_1  \partial_{xy} + 3c_1 v_y  +c_3-\xi_2\partial_x  ) \psi =0,  \label{LaxW1}      \\
&& ( c_2 \partial_{y} + c_1 \partial_{x}^{3} + 3c_1 v_x  \partial_x + \partial_t -\xi_1 +\beta)\psi =0,  \label{LaxW2}
\end{eqnarray}
 where $\xi_1$, $\xi_2$ and $\beta=$ are constants.
Assuming that potential function~$u(x,y)$~tends to~$0$~fast enough  as $|x|+|y|\rightarrow \infty$~, that is to say, the potential function ~$u(x,y)$ belongs to  a Schwartz space $S(R)$
\begin{eqnarray}
S(R)=\{f\in C^{\infty}(R), \|f \|_{\gamma,\beta}=sup |x^\gamma \partial^{\beta}f(x)| < \infty ,\gamma,\beta \in Z_{+} \}  ,     \label{WS}
\end{eqnarray}
then a Jost solution of Eq.(\ref{LaxW1}) satisfies the asymptotic condition
\begin{eqnarray}
\psi(x,y) = e^{izx+(\frac{\xi_2}{3c_1}+\frac{ic_3}{3zc_1})y} +O(1),\,\, |x|,|y|\rightarrow \infty .   \label{pW1}
\end{eqnarray}
Furthermore, resorting to Eq.(\ref{LaxW2}),  the parameter $\beta=ic_3z^3 -c_2(\frac{\xi_2}{3c_1}+\frac{ic_3}{3zc_1})+\xi_1$ is calculated.  For introducing the new Lax pair with $z$ as the spectral parameter, the transformation is taken
\begin{eqnarray}
\phi(x,y,z)=\psi(x,y)e^{-izx-(\frac{\xi_2}{3c_1}+\frac{ic_3}{3zc_1})y}  ,\,\,    \label{pW2}
\end{eqnarray}
an new asymptotic condition of the new parameter has the form
\begin{eqnarray}
\phi(x,y,z)\rightarrow 1 ,\,\, |x|,|y|\rightarrow \infty    .\label{pW3}
\end{eqnarray}
Based on the transformation (\ref{pW2}), the Lax pair~(\ref{LaxW1})-(\ref{LaxW2})~can be changed into
\begin{eqnarray}
&& 3c_1 \phi_{xy}+\frac{ic_3}{z}\phi_x +3izc_1 \phi_y + 3c_1 \partial_x^{-1} u_y \phi=0 , \label{LaxW3}  \\
&&  c_1(\phi_{xxx}+3iz\phi_{xx}-3z^2 \phi_x -iz^3 \phi )+3c_1 v_x (\phi_x +iz\phi) +c_2 [\frac{\xi_2}{3c_1}\phi +\frac{ic_3}{3zc_1}\phi +\phi_y ] +\phi_t-(\xi-\beta)\phi =0
 .\label{LaxW4}
\end{eqnarray}
On the basis of the purpose of searching for a bounded function $\phi(x,y,z)$, the Green's function of Eq.(\ref{LaxW3}) is considered
\begin{eqnarray}
3c_1 G_{xy}+\frac{ic_3}{z} G_x +3izc_1 G_y =\delta(x)\delta(y) . \label{LaxW5}
\end{eqnarray}
According to the Fourier transformation  and the inverse transformation of the multivariate functions and the properties of the $\delta$ function, direct calculation shows
\begin{eqnarray}
{G}(x,y,z)=\frac{1}{4\pi^2} \int_{-\infty}^{\infty} \int_{-\infty}^{\infty}
\frac{e^{i(\xi x+\eta y)}}{-3c_1 \xi \eta - \frac{c_3}{z} \xi- 3zc_1 \eta }d\xi d\eta    .\label{FG1W}
\end{eqnarray}
 Thus, the general solution of Eq.(\ref{LaxW3}) can be expressed as  the convolution of the Green's function $G(x,y,z)$ and $-3c_1 \partial_x^{-1} u_y \phi$, that is
\begin{eqnarray} \label{GSW}
\phi(x,y,z)            =G(x,y,z)*[-3c_1 \partial_x^{-1} u_y \phi]           .
\end{eqnarray}

For the background of zero potential solution $u(x,y)=0$, there are two linearly independent eigenfunctions
\begin{eqnarray}
M_0=1, N_0=e^{-2iz_1 (x+ \frac{c_3}{3c_1 z_1^2 +3c_1 z_2^2}y)} ,
\end{eqnarray}
where $z=z_1+iz_2$. Hence, the two eigenfunctions corresponding to the general potential are  produced, one of them reads
\begin{eqnarray} \label{MW}
 M(x,y,z)=1+  \int_{-\infty}^{\infty} \! \int_{-\infty}^{\infty}\!
G(x-x^{\prime} \!,\! y-y^{\prime}\!,\! z) [-3c_1 \partial_x^{-1} u_{y^{\prime}} M(x^{\prime},y^{\prime},z) dx^{\prime} dy^{\prime},
\end{eqnarray}
and the other is
\begin{eqnarray} \label{NW}
N(x,y,z)=e^{-2iz_1 (x+ \frac{c_3}{3c_1 z_1^2 +3c_1 z_2^2}y)}  + \int_{-\infty}^{\infty} \int_{-\infty}^{\infty}  G(x-x^{\prime},y-y^{\prime},z) (-3c_1 \partial_x^{-1}  u_{y^{\prime}}  N(x^{\prime},y^{\prime},z) )dx^{\prime} dy^{\prime},
\end{eqnarray}
where
\begin{eqnarray}  \label{GW}
G(x-x^{\prime},y-y^{\prime},z) =\frac{1}{4\pi^2} \int_{-\infty}^{\infty} \int_{-\infty}^{\infty}
\frac{e^{i(\xi(x-x^{\prime})+\eta (y-y^{\prime}))}}{-3c_1 \xi \eta - \frac{c_3}{z} \xi- 3zc_1 \eta }d\xi d\eta   \nonumber\\
=\frac{1}{2\pi} \int_{-\infty}^{\infty} g(\xi,y-y^{\prime},z) e^{i\xi(x-x^{\prime})}d\xi ,
\end{eqnarray}
with
\begin{eqnarray}  \label{gW}
g(\xi,y-y^{\prime},z)=\frac{1}{2\pi }\int_{-\infty}^{\infty} \frac{e^{i \eta (y-y^{\prime})}}{-3c_1 \xi \eta - \frac{c_3}{z} \xi- 3zc_1 \eta } d\eta ,
\end{eqnarray}
the first order pole of Eq.(\ref{gW}) is  $\eta_1 =-\frac{c_3 \xi}{ 3c_1 (z_1^2 +2iz_1z_2 -z_2^2 +z_1 \xi +iz_2 \xi)} $.

Let's make a sufficiently large semicircle  $C_R:\eta=Re^{i\theta}, R>\mid \eta_1 \mid, 0\leq \theta \leq \pi$ in the upper half plane  when $y-y^{\prime}>0$ so that $[-R,R]\cup C_R$ forms a closed curve with the counterclockwise direction. $\eta_1$ is inside the curve if $\eta_1$ is in the upper half plane, on the contrary, the integrand is analytic in the curve if $\eta_1$ is in the lower half plane. The residue theorem gives that
\begin{eqnarray}  \label{ResW}
&&  \frac{1}{2\pi i} \int_{-R}^{R} \frac{e^{i\eta (y-y^{\prime})}}{-3c_1 \xi \eta - \frac{c_3}{z} \xi- 3zc_1 \eta }d\eta+\frac{1}{2\pi i} \int_{C_R} \frac{e^{i\eta (y-y^{\prime})}}{-3c_1 \xi \eta - \frac{c_3}{z} \xi- 3zc_1 \eta }d\eta    \nonumber\\
&&  =\underset{\eta=\eta_1,Im\eta_1 > 0} {Res}[\frac{e^{i\eta (y-y^{\prime})}}{-3c_1 \xi \eta - \frac{c_3}{z} \xi- 3zc_1 \eta}] .
\end{eqnarray}
When $R\rightarrow \infty$,  the limit of  the second integral on the left-hand side is 0 due to the Jordan theorem, therefore, we have
\begin{eqnarray}  \label{g1W}
&&  g(\xi,y-y^{\prime},z)=\underset{\eta=\eta_1} {Res}[\frac{e^{i\eta (y-y^{\prime})}}
    {-3c_1 \xi \eta - \frac{c_3}{z} \xi- 3zc_1 \eta }] \nonumber \\
&&   =
\begin{cases}
 e^{-i \frac{c_3 \xi \cdot [z_1^2-z_2^2 +z_1 \xi -i(2z_1z_2+z_2 \xi)]}{3c_1 [(z_1^2-z_2^2+z_1\xi)^2 + (2z_1z_2 +z_2 \xi)^2 ] } (y-y^{\prime}) } ,
 \frac{c_3 \xi (2z_1z_2 +z_2 \xi)}{ 3c_1[(z_1^2-z_2^2 +z_1 \xi)^2 + (2z_1z_2 +z_2\xi)^2]} > 0 , \\
0   , \quad\quad\quad \quad \quad\quad\quad\quad\quad\quad\quad\quad\quad\quad    \frac{c_3 \xi (2z_1z_2 +z_2 \xi)}{ 3c_1[(z_1^2-z_2^2 +z_1 \xi)^2 + (2z_1z_2 +z_2\xi)^2]}  < 0  ,
\end{cases}      \nonumber \\
&& =H[ \frac{c_3 \xi (2z_1z_2 +z_2 \xi)}{ 3c_1[(z_1^2-z_2^2 +z_1 \xi)^2 + (2z_1z_2 +z_2\xi)^2]} (y-y^{\prime})  ]
e^{-i \frac{c_3 \xi \cdot [z_1^2-z_2^2 +z_1 \xi -i(2z_1z_2+z_2 \xi)]}{3c_1 [(z_1^2-z_2^2+z_1\xi)^2 + (2z_1z_2 +z_2 \xi)^2 ] } (y-y^{\prime}) }.
\end{eqnarray}
When $y-y^{\prime}<0$,  making a sufficiently large semicircle $C_R:\eta=Re^{i\theta}, -\pi \leq \theta \leq 0$  analogously in the lower half plane yields
\begin{eqnarray}  \label{g2W}
 &&  g(\xi,y-y^{\prime},z)=
\begin{cases}
 -e^{-i \frac{c_3 \xi \cdot [z_1^2-z_2^2 +z_1 \xi -i(2z_1z_2+z_2 \xi)]}{3c_1 [(z_1^2-z_2^2+z_1\xi)^2 + (2z_1z_2 +z_2 \xi)^2 ] } (y-y^{\prime}) } ,
 \frac{c_3 \xi (2z_1z_2 +z_2 \xi)}{ 3c_1[(z_1^2-z_2^2 +z_1 \xi)^2 + (2z_1z_2 +z_2\xi)^2]}< 0 , \\
0   , \quad\quad\quad \quad \quad \quad\quad\quad\quad\quad\quad\quad\quad\quad\quad     \frac{c_3 \xi (2z_1z_2 +z_2 \xi)}{ 3c_1[(z_1^2-z_2^2 +z_1 \xi)^2 + (2z_1z_2 +z_2\xi)^2]}  > 0  ,
\end{cases}      \nonumber \\
&& = -H[  \frac{c_3 \xi (2z_1z_2 +z_2 \xi)}{ 3c_1[(z_1^2-z_2^2 +z_1 \xi)^2 + (2z_1z_2 +z_2\xi)^2]} (y-y^{\prime})]
e^{-i \frac{c_3 \xi \cdot [z_1^2-z_2^2 +z_1 \xi -i(2z_1z_2+z_2 \xi)]}{3c_1 [(z_1^2-z_2^2+z_1\xi)^2 + (2z_1z_2 +z_2 \xi)^2 ] } (y-y^{\prime}) } .
\end{eqnarray}
Combining the two formulas (\ref{g1W}) and (\ref{g2W}), the Green's function is
\begin{eqnarray}  \label{GgW}
&& G(x-x^{\prime},y-y^{\prime},z) =\frac{sgn(y-y^{\prime})}{2\pi} \int_{-\infty}^{\infty}    H[\frac{c_3 \xi (2z_1z_2 +z_2 \xi)}{ 3c_1[(z_1^2-z_2^2 +z_1 \xi)^2 + (2z_1z_2 +z_2\xi)^2]} (y-y^{\prime}) ]  \nonumber\\
&& \cdot  e^{-i \frac{c_3 \xi \cdot [z_1^2-z_2^2 +z_1 \xi -i(2z_1z_2+z_2 \xi)]}{3c_1 [(z_1^2-z_2^2+z_1\xi)^2 + (2z_1z_2 +z_2 \xi)^2 ] } (y-y^{\prime}) +i\xi(x-x^{\prime})}d\xi .
\end{eqnarray}
%Similarly, Ablowitz et al. found that the $\bar{\partial}$  problem can be employed to solve the inverse scattering equation in cases where the eigenfunctions are not analytic to formulate inverse scattering equations~\cite{Ablowitz1991,ablowitz1983Dbar}.
Because the Green's function has no jump across the real axis and this function is analytic in the $z$-plane. While the eigenfunctions $M,N$ are not analytic in the same plane, the Riemann-Hilbert method can not be applied. That is why we use the $\bar{\partial}$ method to study solutions of Eq.(\ref{Wave}).

\subsection{scattering equation and $\bar{\partial}$ problem }
Firstly,  the $\bar{\partial}$ derivative is calculated as follows:
\begin{eqnarray}  \label{dbarM}
&& \bar{\partial}M(x,y,z)=  \int_{-\infty}^{\infty} \int_{-\infty}^{\infty}[\bar{\partial}G(x-x^{\prime},y-y^{\prime},z)][-3c_1 \partial_{x}^{-1} u_{y^{\prime}}(x^{\prime},y^{\prime})M(x^{\prime},y^{\prime},z) ]dx^{\prime} dy^{\prime}          \nonumber    \\
&& +\int_{-\infty}^{\infty} \int_{-\infty}^{\infty}G(x-x^{\prime},y-y^{\prime},z)[ -3c_1 \partial_{x}^{-1} u_{y^{\prime}}(x^{\prime},y^{\prime}) \bar{\partial}M(x^{\prime},y^{\prime},z) ]dx^{\prime} dy^{\prime}\! ,
\end{eqnarray}
where
\begin{eqnarray}  \label{dbarG}
&& \bar{\partial}G(x-x^{\prime},y-y^{\prime},z)=\frac{1}{4\pi^2} \int_{-\infty}^{\infty} \int_{-\infty}^{\infty}
e^{i(\xi(x-x^{\prime})+\eta (y-y^{\prime}))}\bar{\partial}{\frac{1}{-3c_1 \xi \eta - \frac{c_3}{z} \xi- 3zc_1 \eta }}d\xi d\eta  ,  \nonumber\\
&&=
\begin{cases}
- \frac{i}{12c_1 \pi}  \int_{-\infty}^{\infty} \int_{-\infty}^{\infty}
 \frac{z  e^{i(\xi(x-x^{\prime})+\eta (y-y^{\prime}))} }{\mid z_2 \mid \mid z_1^2-z_2^2+z_1\xi \mid}
\delta(\eta+\frac{c_3 \xi}{3c_1(z_1^2-z_2^2 +z_1\xi)}  )\delta(\xi+2z_1)    d\xi d\eta ,  \quad  \eta > 0  , \\
\frac{i}{12c_1 \pi}  \int_{-\infty}^{\infty} \int_{-\infty}^{\infty}
 \frac{z  e^{i(\xi(x-x^{\prime})+\eta (y-y^{\prime}))} }{\mid z_2 \mid \mid z_1^2-z_2^2+z_1\xi \mid}
\delta(\eta+\frac{c_3 \xi}{3c_1(z_1^2-z_2^2 +z_1\xi)}  )\delta(\xi+2z_1)    d\xi d\eta ,  \quad\quad  \eta < 0  ,\nonumber\\
     \end{cases}      \nonumber \\
&& =
\begin{cases}
-\frac{1}{12c_1 \pi}   \frac{z_1+iz_2}{\mid z_2 \mid \mid z_1^2-z_2^2+z_1\xi \mid}    e^{-2iz_1(x-x^{\prime})- i \frac{2z_1c_3}{3c_1(z_1^2+z_2^2)} (y-y^{\prime})},  \quad  c_3c_1< 0,  \\
\frac{1}{12c_1 \pi}   \frac{z_1+iz_2}{\mid z_2 \mid \mid z_1^2-z_2^2+z_1\xi \mid}    e^{-2iz_1(x-x^{\prime})- i \frac{2z_1c_3}{3c_1(z_1^2+z_2^2)} (y-y^{\prime})},  \quad\quad  c_3c_1 > 0, \nonumber\\
\end{cases}
\end{eqnarray}
which implies  the following equation
\begin{eqnarray}  \label{dbarM1W}
&& \bar{\partial}M(x,y,z)=F(z_1,z_2)  e^{-2iz_1[x+\frac{c_3}{3c_1(z_1^2+z_2^2)}y]}  \nonumber\\
&&  +\int_{-\infty}^{\infty} \int_{-\infty}^{\infty}G(x-x^{\prime},y-y^{\prime},z)[-3c_1 \partial_x^{-1}u_{y^{\prime}}(x^{\prime},y^{\prime})\bar{\partial}M (x^{\prime},y^{\prime},z)]dx^{\prime} dy^{\prime},
\end{eqnarray}
where  the spectral data $F(z_1,z_2)$ is
\begin{eqnarray}  \label{FW}
F(z_1,z_2)=\begin{cases}
\frac{1}{12c_1 \pi} \frac{z_1+iz_2}{(z_1^2+z_2^2)\mid z_2 \mid }  \int_{-\infty}^{\infty} \int_{-\infty}^{\infty}
3c_1 \partial_{x}^{-1} u_{y^{\prime}}(x^{\prime},y^{\prime}) M(x^{\prime},y^{\prime},z) e^{2iz_1[x^{\prime} +\frac{c_3}{3c_1(z_1^2+z_2^2)}y^{\prime} ]}dx^{\prime}dy^{\prime},  \quad c_3c_1< 0 , \\
-\frac{1}{12c_1 \pi} \frac{z_1+iz_2}{(z_1^2+z_2^2)\mid z_2 \mid }  \int_{-\infty}^{\infty} \int_{-\infty}^{\infty}
3c_1 \partial_{x}^{-1} u_{y^{\prime}}(x^{\prime},y^{\prime}) M(x^{\prime},y^{\prime},z) e^{2iz_1[x^{\prime} +\frac{c_3}{3c_1(z_1^2+z_2^2)}y^{\prime} ]}dx^{\prime}dy^{\prime},  \,\,  c_3c_1 > 0 .
\end{cases}
\end{eqnarray}

Let's multiply both sides of Eq.(\ref{NW}) by $F(z_1,z_2)$  and subtract Eq.(\ref{dbarM1W}) and suppose that the corresponding homogeneous
integral equation has only zero solutions, the  scattering equation of $\bar{\partial}$ problem is received by comparing Eq.(\ref{dbarM1W}) with the known Eq.(\ref{NW})
\begin{eqnarray}  \label{FN-dbarMW}
\bar{\partial}M(x,y,z)= F(z_1,z_2)N(x,y,z).
\end{eqnarray}

We are committed to a purpose to look for  a symmetry relation  of Green's function so that $N$ can be represented by $M$, then
\begin{eqnarray}  \label{GzW}
&& G(x-x^{\prime},y-y^{\prime},-\bar{z}) =\frac{1}{4\pi^2} \int_{-\infty}^{\infty} \int_{-\infty}^{\infty}
\frac{e^{i(\xi(x-x^{\prime})+\eta (y-y^{\prime}))}}{-3c_1 \xi \eta + \frac{c_3}{\bar{z}} \xi + 3\bar{z} c_1 \eta }d\xi d\eta   \nonumber\\
&& =\frac{1}{4\pi^2} \int_{-\infty}^{\infty} \int_{-\infty}^{\infty}
\frac{e^{i(\xi(x-x^{\prime})+\eta (y-y^{\prime}))}}{-3c_1(\xi-2z_1)(\eta-\frac{2z_1c_3}{3c_1(z_1^2+z_2^2)})  -\frac{c_3}{z_1+iz_2}(\xi-2z_1) -3(z_1+iz_2)c_1 (\eta-\frac{2z_1c_3}{3c_1(z_1^2+z_2^2)})} d\xi d\eta ,   \nonumber\\
\end{eqnarray}
by taking the transformation $\xi-2z_1 \rightarrow \xi$, $\eta-\frac{2z_1c_3}{3c_1(z_1^2+z_2^2)} \rightarrow \eta$, the symmetry property of the Green's function obeys
\begin{eqnarray}  \label{Gz1W}
G(x-x^{\prime},y-y^{\prime},-\bar{z}) =G(x-x^{\prime},y-y^{\prime},z) e^{2iz_1[(x-x^{\prime})+\frac{c_3}{3c_1(z_1^2+z_2^2)}(y-y^{\prime})]}  .
\end{eqnarray}
Replacing $z$ with $-\bar{z}$ in Eq.(\ref{MW}) and considering that
\begin{eqnarray}  \label{M-zW}
&& M(x,y,-\bar{z}) e^{-2iz_1[x+\frac{c_3}{3c_1(z_1^2+z_2^2)} y]}=e^{-2iz_1[x + \frac{c_3}{3c_1(z_1^2+z_2^2)}y]}     \nonumber\\
&& +\int_{-\infty}^{\infty} \int_{-\infty}^{\infty}
 G(x-x^{\prime},y-y^{\prime},z)[-3c_1 \partial_x^{-1}u_{y^{\prime}}(x^{\prime},y^{\prime}) M (x^{\prime},y^{\prime},z) e^{-2iz_1(x^{\prime} +\frac{c_3}{3c_1(z_1^2+z_2^2)}y^{\prime})} ] dx^{\prime} dy^{\prime},
\end{eqnarray}
by comparing with the eigenfunction $N$, the other relation can be described as
\begin{eqnarray}  \label{MNW}
N(x,y,z)=M(x,y,-\bar{z}) e^{-2iz_1[x+\frac{c_3}{3c_1(z_1^2+z_2^2)} y]} .
\end{eqnarray}
Hence, the $\bar{\partial}$ problem for the eigenfunctions $M$ and $N$ is further expressed  by substituting (\ref{MNW}) into (\ref{FN-dbarMW})  in terms of the scattering data:
\begin{eqnarray}  \label{MNdbar}
\bar{\partial}M(x,y,z)=F(z_1,z_2)M(x,y,-\bar{z}) e^{-2iz_1[x+\frac{c_3}{3c_1(z_1^2+z_2^2)} y]} .
\end{eqnarray}
%It should be stressed that the above calculation is valid under the assumption that there are no nontrivial solutions of the homogeneous integral equation obtained from Eq.(\ref{N}).
\subsection{Inverse spectral problem}

The inverse spectral problem can be solved by using the Cauchy-Green formula
\begin{eqnarray}  \label{MsolW}
M(x,y,z)=1+\frac{1}{2\pi i} \int \int \frac{d\zeta \wedge d\bar{\zeta}}{\zeta-z}
F(\zeta_1,\zeta_2)M(x,y,-\bar{\zeta}) e^{-2i\zeta_1[x+ \frac{c_3}{3c_1(z_1^2+z_2^2)}y]} ,
\end{eqnarray}
%By the way, we can write the solution
%\begin{eqnarray}  \label{Mxsol}
%M(x,y,z)+izM(x,y,z)=1+\frac{1}{2\pi i} \int \int \frac{d\zeta \wedge d\bar{\zeta}}{\zeta-z}
%F(\zeta_1,\zeta_2)[M_x(x,y,\zeta)+izM(x,y,\zeta)] e^{-2i\zeta_1[x+(\zeta_1^2-3\zeta_2^2)y]}
%\end{eqnarray}
where $\zeta=\zeta_1+i \zeta_2$.
In order to reconstruct the potential function $u(x,y)$, it is significant to compare the two expressions of $M-1$,
\begin{eqnarray}  \label{doubleM-1W}
&&  M(x,y,z)-1=
\int_{-\infty}^{\infty} \int_{-\infty}^{\infty}
G(x-x^{\prime},y-y^{\prime},z) [-3c_1 \partial_x^{-1}u_{y^{\prime}}(x^{\prime},y^{\prime}) M (x^{\prime},y^{\prime},z)]
dx^{\prime} dy^{\prime},    \nonumber \\
&& M(x,y,z)-1=  \frac{1}{2\pi i} \int_{-\infty}^{\infty} \int_{-\infty}^{\infty} \frac{d\zeta\wedge d\bar{\zeta}} {\zeta-z} F(\zeta_1,\zeta_2)
M(x,y,-\bar{\zeta} )e^{-2i\zeta_1(x + \frac{c_3}{3c_1(z_1^2+z_2^2)} y)}.
\end{eqnarray}

In what follows, the potential function $u(x,y)$ is established by comparing the order of $O(z^{-1})$, resort to the Green's function for $\mid z \mid\rightarrow \infty$,   the asymptotic expansion of Green's function is
\begin{eqnarray}  \label{GoW}
&& G(x,y,z,\bar{z}) =\frac{1}{4\pi^2} \int_{-\infty}^{\infty} \int_{-\infty}^{\infty}
   \frac{e^{i(\xi x+\eta y)}}{-3c_1 \xi \eta - \frac{c_3}{z} \xi- 3zc_1 \eta  }d\xi d\eta   \nonumber\\
&&    =- \frac{1}{12 \pi ^2 c_1 z} \int_{-\infty}^{\infty} \frac{e^{i\eta y}} {\eta} d\eta \int_{-\infty}^{\infty} e^{i\xi x} d\xi+O(z^{-2})        \nonumber\\
&&  = -\frac{i}{6c_1 z}sgn(y)\delta(x) +O(z^{-2}) .
\end{eqnarray}
Substituting Eq.(\ref{GoW}) into the first one of Eq.(\ref{doubleM-1W}),   the asymptotic expansion can be derived as
\begin{eqnarray}  \label{M-1W}
M(x,y,z)-1 = \frac{i}{z}\partial_x^{-1} u(x,y)   +O(z^{-2}),
\end{eqnarray}
the asymptotic expression of second equation (\ref{doubleM-1W})  from the $\bar{\partial}$ representation is
\begin{eqnarray}  \label{M-12W}
M(x,y,z)-1 =  \frac{i}{2\pi z}\int\int  F(\zeta_1,\zeta_2)M(x,y,-\bar{\zeta})e^{-2i\zeta_1[x+\frac{c_3}{3c_1(z_1^2+z_2^2)}y]}  d\zeta \wedge d\bar{\zeta}   +O(z^{-2}) .
\end{eqnarray}
Comparing Eq.(\ref{M-1W}) and Eq.(\ref{M-12W}), we finally arrived at the reconstruction formula of the potential function $u(x,y)$,
\begin{eqnarray}  \label{uW}
u(x,y) = \frac{1}{2\pi } \partial_x \int\int d\xi \wedge d\bar{\xi} F(\xi_1,\xi_2) M(x,y,-\bar{\xi}) e^{-2i\xi_1 (x+\frac{c_3}{3c_1(z_1^2+z_2^2)} y)}.
\end{eqnarray}
A solution of (2+1)-dimensional nonlinear wave equation is acquired if the time evolution of spectral data is determined. Hence, inserting Eq.(\ref{pW2}) into $\bar{\partial}$-equation
\begin{eqnarray}  \label{psidbarW}
\bar{\partial}\psi(x,y,t,z)=F(z_1,z_2,t)\psi(x,y,t,-\bar{z}) ,
\end{eqnarray}
we take the derivative of both sides of this equation in regard to $t$
\begin{eqnarray}  \label{psitW}
[\bar{\partial}\psi(z)]_t=F_t(z_1,z_2,t)\psi(-\bar{z})+F(z_1,z_2,t)\psi_t(-\bar{z}) .
\end{eqnarray}
The parameter $\beta=ic_3z^3 -c_2(\frac{\xi_2}{3c_1}+\frac{ic_3}{3zc_1})+\xi_1$ is given by observing Eq.(\ref{LaxW2}) which $\psi(z)$ is satisfied and $u\in S(R)$, which leads to the following result in touch with Eq.(\ref{psitW})
\begin{eqnarray}  \label{F1W}
F(z_1,z_2,t)=F(z_1,z_2)e^{(-ic_3z^3-ic_3\bar{z}^3 +\frac{ic_3}{3zc_1} +\frac{ic_3}{3\bar{z} c_1})   t} ,
\end{eqnarray}
where $F(z_1,z_2,0)=F(z_1,z_2)$.
Therefore, the solution (\ref{uW})  can be represented as
\begin{eqnarray}  \label{uWt}
u(x,y,t) = \frac{1}{2\pi } \partial_x \int\int d\xi \wedge d\bar{\xi} F(\xi_1,\xi_2) M(x,y,t,-\bar{\xi}) e^{-2i\xi_1 (x+\frac{c_3}{3c_1(z_1^2+z_2^2)} y)+(-ic_3z^3-ic_3\bar{z}^3 +\frac{ic_3}{3zc_1} +\frac{ic_3}{3\bar{z} c_1} )t}.
\end{eqnarray}

\section{conclusion}
In the paper, based on the spectra problem, a novel systematical solution method of the (2+1)-dimensional Kaup-Kuperschmidt equation and a generalized (2+1)-dimensional nonlinear wave equation are studied  by means of the $\bar{\partial}$-dressing method. To this end, we bring in the eigenfunctions and Green's functions of the (2+1)-dimensional equations  firstly,
specially noting that the eigenfunctions and Green's functions play an important role in receiving their solutions. According to the Fourier transformation and inverse Fourier transformation, the Green's functions can be confirmed. By analyzing the Green's functions, we conclude that they didn't have jump across the real axis but they are analytic in the $z$-plane. However, the eigenfunctions were not analytic, so the non-analyticity of which prevents  from making using of the Riemann-Hilbert problem. Fortunately, the $\bar{\partial}$ problem  is found to be useful for formulating scattering equations and constructing soliton solutions. By constructing the $\bar{\partial}$ equation and solving the inverse spectrum problem of the (2+1)-dimensional equations, the soliton solutions are obtained if the time evolutions are determined.

The reduced problem of nonlinear equation has always been an important topic in the field of nonlinear science. The main method is binary Bell polynomials related about the $\mathcal{Y}$-constraints and $P$-conditions which turns out to be appropriate tools to represent the bilinear form. They are in one to one correspondence with the more familiar bilinear expressions of Hirota's operator, known to facilitate the search of the NLEE's explicit solutions. With the help of this method, under the scaling transformation, the bilinear form is determined.
Starting from the bilinear representation, the CKP equation is  reduced to Kaup-Kuperschmidt~equation and bidirectional Kaup-Kuperschmidt equation under the reduction of $t_3$ and $t_5$, respectively. I believe that there are still many deep relations between generalized Bell polynomials and integrable structures,  which still remain open and worth to be studied including the Miura transformation of the reduced equations, symmetry, etc. \\

%展望一下AKP，BKP，CKP的约化后的方程之间的联系，例如miura变换等的。以及再做个桥梁联系Dbar和贝尔多项式的联系。

{\bf ACKNOWLEDGEMENT } This work was supported by the National Natural Science Foundation of China (grant No.11971475). \\
%{\bf Acknowledgement } This work was supported by the National Natural Science Foundation of China (grant No.11971475). \\

%{\bf AUTHORS' CONTRIBUTIONS } Xuedong Chai and Yufeng Zhang contributed equally to this work. \\

{\bf CONFLICT OF INTEREST}: The authors declare that they have no conflicts of interest.\\

 {\bf DATA AVAILABILITY}: The data that support the findings of this study are available within the article.

{\bf Acknowledgement } This work was supported by the National Natural Science Foundation of China (grant No.11971475, No.12001246) and the NSF of Jiangsu Province of China (Grant No. BK20190991) and the NSF of Jiangsu Higher Education Institutions of China (Grant No. 19KJB110011)\\

\end{document}